\documentclass[aps,prb,twocolumn]{revtex4}
\usepackage{graphicx}
\def\bea{\begin{eqnarray}}
\def\eea{\end{eqnarray}}
\def\ben{\begin{equation}}
\def\een{\end{equation}}
\def\benu{\begin{enumerate}}
\def\enu{\end{enumerate}}
\def\scr{\scriptstyle}

\def\ps{\psi}
\def\psd{\psi^{\dagger}}
\def\dps{\dot{\psi}}
\def\dpsd{\dot{\psi}^{\dagger}}
\def\cd{\dot{c}}
\def\nd{\dot{n}}
\def\hd{\dot{h}}

\def\da{\dagger}
\def\Ws{W^{\ast}}
\def\vv{{\bf v}}
\def\vr{{\bf r}}
\def\vrp{{\bf r}^{\prime}}
\def\rpm{r^{\prime}}
\def\vR{{\bf R}}
\def\vj{{\bf j}}
\def\vJ{{\bf J}}
\def\vpot{{\bf A}}
\def\pol{{\bf P}_{{\rm el}}}
\def\vp{{\bf p}}
\def\vq{{\bf q}}
\def\vk{{\bf k}}
\def\vkp{{\bf k}^{\prime}}
\def\vL{{\bf L}}

\def\ham{{\mathcal H}}
\def\mg{{\mathcal G}}
\def\ord{{\mathcal O}}
\def\pno{{\mathcal N}}
\def\op{\hat{O}}
\def\grad{{\bf \nabla}}
\def\prl{\partial}
\def\dl{\delta}
\def\ph{\phi}
\def\rh{\rho}
\def\ga{\gamma}
\def\si{\sigma}
\def\ep{\epsilon}
\def\om{\omega}
\def\Om{\Omega}
\def\ta{\tau}
\def\al{\alpha}
\def\be{\beta}
\def\sip{\sigma^{\prime}}
\def\sib{\bar{\sigma}}
\def\la{\langle}
\def\ra{\rangle}
\begin{document}

\title{Thermal Transport for Many Body Tight-Binding Models}

\author{Indranil Paul and Gabriel Kotliar}

\affiliation{Center for Materials Theory, Department of Physics and Astronomy,
Rutgers University, Piscataway, New Jersey 08854}
\date{\today}

\begin{abstract}
We clarify some aspects of the calculation of the thermal transport 
coefficients.
For a tight-binding Hamiltonian we discuss the approximate nature of the
charge current and the thermal current obtained by Peierls substitution
which is also identical to the equation of motion technique. We address the 
issue
of choosing an appropriate basis for making the Peierls construction for 
transport calculations. 
We propose a criteria for finding an
optimum Wannier basis where the difference between the exact current and 
the approximate one is minimum. Using the equations of motion we derive 
the thermal current for a generalized Hubbard model with density 
interaction. We identify a part which is the contribution from the long range
interactions to the heat current. For the Hubbard model we derive expressions
for the transport coefficients in the limit of infinite dimensions.
\end{abstract}

\pacs{PACS numbers: 72.15.Eb, 71.10.Fd, 72.15.Jf, 72.20.Pa}

\maketitle

\section{Introduction}
The theoretical description of the thermoelectric response of correlated 
materials is a fundamental problem in condensed matter physics, and a 
breakthrough in this area has potential technological useful 
implications.~\cite{0mahan} 
The materials, which have been studied as likely candidates for useful
thermoelectric properties, are mostly semiconductor alloys and compounds.
Materials such as Bi$_2$Te$_3$/Sb$_2$Te$_3$ and Si-Ge, which are currently 
favoured for room temperature application, belong to this category. Another
class of materials, with potentially useful thermoelectric properties, are 
Ce and La filled skutterudites such as LaFe$_3$CoSb$_{12}$ and 
CeFe$_3$CoSb$_{12}$.~\cite{0mahan} Theoretically these materials have been 
studied successfully using band theory.~\cite{dsingh} Recently Mahan and 
Sofo~\cite{sofo} have shown that the best thermoelectric materials could well
be correlated metals and semiconductors (i.e., rare earth intermetallic 
compounds).
The development of the dynamical mean field theory
(DMFT)[for reviews see Refs. 4, 5] has allowed new studies of the effects of 
correlation on the thermoelectric response using this method on model
Hamiltonians.~\cite{gunnar,udo,freericks} More recent combinations of band 
theory and many-body methods such as
the LDA+DMFT method~\cite{anisimov} [for reviews see Refs. 
10, 11] or the LDA++ method~\cite{lichtenstein} offers the exciting 
possibility of predicting the thermoelectric 
properties of materials starting from first principles.~\cite{2gunnar} This 
revival of interest in the 
thermoelectric response motivates us to re-analyze in this paper the following
issues: (1) what is the form of the thermal current and the charge current 
which should be used in
realistic calculations, and (2) how it should be approximated in a DMFT 
calculation.

The first question is subtle for two reasons. First, as noted early on by 
Jonson and Mahan,~\cite{jonson} the electronic part of the thermal current 
operator contains a 
quadratic and a quartic piece (if the electron-electron interaction is 
non-local) in the electron creation and annihilation operators. The 
contribution of this quartic interaction term to the current has continued to 
be the 
subject of discussion.~\cite{moreno} Second, while the form of the thermal 
current and the charge current in the continuum is unambiguous, and can be 
calculated using Noether's theorem,~\cite{langer,ryder} DMFT calculations 
require the projection of these currents on a restricted lattice model. This 
involves the computation of complicated matrix elements, and in practice 
an approximation which is analogous to the Peierls substitution~\cite{peierls}
for the 
electrical current is carried out. It is well known that the results of this
construction depend on the basis set of orbitals used.~\cite{millis}
This raises the 
practical question of how to optimize the basis of orbitals to be used in 
transport calculations. 

The second question is subtle due to the presence
of interaction terms in the current. This raises the issue of how it should 
be simplified in the evaluation of the various current-current correlation 
functions and the transport coefficients.
This question was first addressed by Schweitzer and Czycholl~\cite{schweitzer}
and by Pruschke and collaborators~\cite{pruschke}
who stated that \emph{within the relaxation time approximation},
this term can be expressed in terms of a time derivative, and the vertex 
corrections can be ignored. In the review of Georges 
\emph{et. al.}~\cite{gabi} it was stated that the results of Pruschke
\emph{et. al.} hold beyond the relaxation time approximation in the limit of 
large dimensionality when DMFT becomes exact but no detailed proof of this 
statement was presented.

The following are our main results. (1) In section II we address the question
of the optimization of the basis of localized orbitals for transport 
calculations, following the ideas of Marzari and Vanderbilt.~\cite{marzari}
For completeness and for pedagogical reasons we discuss in 
parallel work on the charge current, which is simpler and better 
understood~\cite{blount} than the thermal current. Our conclusions in this 
context have applications
for the computation of Born charges in empirical tight-binding models.~\cite
{bennetto} (2) In section III we derive the form of the thermal current to be 
used in tight-binding models, and its dependence on the orbitals, using the 
equation of motion technique introduced in Ref. 24. Our final expression
differs in one term from the results of Ref. 15. (3) In section IV we 
describe in detail the diagrammatic analysis of correlation functions of the 
current operators. We demonstrate explicitly that in the DMFT limit of the 
transport calculation, the vertex corrections (even for those involving the
thermal current) can be completely neglected, thereby
justifying the current practice used in all previous DMFT work.  

\section{Charge Current}
We consider a system of electrons in a periodic potential $V(r)$, in the 
presence of an external vector potential $\vpot(r)$, and 
with coulomb interaction between them. The Lagrangian is given by
\begin{widetext}
\bea
L = 
&& 
\frac{i}{2} \int d^3 r \left( \psd \dps - \dpsd \ps \right) + 
\frac{1}{2m} \int d^3 r \psd \left( \grad - ie \vpot(r) \right)^2 \ps 
- \int d^3 r V(r) \psd \ps 
\nonumber \\ && 
- \frac{e^2}{2} 
\int \int d^3r d^3 \rpm \psd(\vr) \psd(\vrp) \frac{1}{\left| \vr - \vrp 
\right|} \ps(\vrp) \ps(\vr).
\eea
\end{widetext}
Here $\psd(\vr)$ and $\ps(\vr)$ are the electron field operators with usual 
anticommutation properties. 
We have ignored the spin of the electrons only to simplify the notation. 
Including spin in the following analysis is quite straightforward.
In field theory, when both high and low energy 
degrees of freedom are retained, Noether's theorem provides a 
robust procedure to identify the various currents.~\cite{ryder} The theorem 
associates with every symmetry of the action a conserved charge and a 
corresponding current. The charge current is determined by the invariance of 
the action $S = \int dt L(t)$, under $U(1)$ gauge transformation given by
$ \ps(\vr) \rightarrow \ps(\vr) e^{i \ph(\vr)}$ and $ \psd(\vr) \rightarrow 
\psd(\vr) e^{-i \ph(\vr)} $. The transformation does not produce any variation
from the interaction term, and the well known expression for the charge 
current is
\ben
\vj = - \frac{ie}{m} \int d^3r \psd (\vr) \left( \grad - ie \vpot(r) \right)
\ps (\vr).
\een
The above expression is gauge invariant. The part which is proportional to the
vector potential gives the diamagnetic current.

In order to facilitate further discussion we will perform the standard Noether
construction in the Wannier basis. In this basis the
action (which includes both low and high energy degrees of freedom) is
\begin{widetext}
\bea
S = \int dt && \left\{ \frac{i}{2} \sum_{n \mu} \left( c_n^{\da \mu} 
\cd_n^{\mu} - \cd_n^{\da \mu} c_n^{\mu} \right) -\sum_{\scr n m \atop 
\scr \mu \nu} t_{n m}^{\mu \nu} c_n^{\da \mu} c_m^{\nu} 
+ \frac{e}{2 m} \sum_{\scr n m l \atop \scr \mu \nu \ga} \vp_{n l}^{\mu \ga}
\cdot \vpot_{l m}^{\ga \nu} c_n^{\da \mu} c_m^{\nu} 
+ \frac{e}{2 m} \sum_{\scr n m l \atop \scr \mu \nu \ga} \vpot_{n l}^{\mu \ga}
\cdot \vp_{l m}^{\ga \nu} c_n^{\da \mu} c_m^{\nu}
\right.
\nonumber \\
&&
\left.
- \frac{e^2}{2 m} \sum_{\scr n m l
\atop \scr \mu \nu \ga} \vpot_{n l}^{\mu \ga} \cdot \vpot_{l m}^{\ga \nu}
c_n^{\da \mu} c_m^{\nu}
- \frac{1}{2}
\sum_{\scr n_1 \ldots n_4 \atop \scr \mu_1 \ldots \mu_4} U_{n_1 \ldots n_4}
^{\mu_1 \ldots \mu_4} c_{n_1}^{\da \mu_1} c_{n_2}^{\da \mu_2} c_{n_3}^{\mu_3}
c_{n_4}^{\mu_4} \right\},
\eea
\end{widetext}
where $t_{n m}^{\mu \nu} = \la n \mu | \ham_0 | m \nu \ra$, 
$\vp_{n m}^{\mu \nu} = \la n \mu | \vp | m \nu \ra$, $\vpot_{n m}^{\mu \nu}
=\la n \mu | \vpot(r) | m \nu \ra$,
and 
$U_{n_1 \ldots n_4}^{\mu_1 \ldots \mu_4} = \la n_1 \mu_1, n_2 \mu_2 | e^2/
| \vr - \vrp | | n_4 \mu_4, n_3 \mu_3 \ra$. Here $\ham_0 = \vp^2/2m + V(r)$ 
is the non-interacting part of the Hamiltonian, $\mu$ is the band index, and 
$\vR_n$ defines the lattice positions. $W_{\mu} (\vr - \vR_n) = \la \vr | n 
\mu \ra $ form a complete set of orthonormal Wannier functions. The creation 
and annihilation operators satisfy the anticommutation relation $\{ c_n^{\mu},
c_m^{\da \nu} \} = \dl_{n m} \dl_{\mu \nu}$. The gauge transformation of
the fermionic field operators is equivalent to the variation $\dl c_n^{\mu}
= i \int d^3r \ph(\vr) \ps(\vr) \Ws_{\mu} (\vr - \vR_n)$ and 
$\dl c_n^{\da \mu} = -i \int d^3r \ph(\vr) \psd(\vr) W_{\mu} (\vr - \vR_n)$.
Expanding $\ph(\vr)$ about the point $\vR_n$ and keeping only up to $\grad
\ph $ term (which is all we need to construct the Noether current) we get
\bea
\dl c_n^{\mu} &=& i \ph(\vR_n) c_n^{\mu} + i \grad \ph \sum_{m \nu} \vL_{nm}
^{\mu \nu} c_m^{\nu},
\nonumber \\
\dl c_n^{\da \mu} &=& -i \ph(\vR_n) c_n^{\da \mu} - i \grad \ph 
\sum_{m \nu} c_m^{\da \nu} \vL_{mn}^{\nu \mu},
\eea
where $\vL_{nm}^{\mu \nu} = \int d^3 r \Ws_{\mu} (\vr - \vR_n) (\vr - \vR_n) 
W_{\nu}(\vr - \vR_m)$ are the connection coefficients. The matrix $\vL$ is 
hermitian, i.e., ${\vL^{\ast}}_{nm}^{\mu \nu} = \vL_{mn}^{\nu \mu}$. 
We note first that the variation from the interaction term is 
exactly zero. Next, using the operator identity $[r_i, A_j(r)] = 0$ we find 
that the variation from the term quadratic in $\vpot(r)$ is zero. To get the 
correct diamagnetic part we make use of $[r_i, p_j] = i \dl_{ij}$.
From the invariance of the action we can identify the charge current as
\begin{widetext}
\bea
\vj &=& ie \sum_{\scr nm \atop \scr \mu \nu}  
(\vR_m - \vR_n ) t_{nm}^{\mu \nu} 
c_n^{\da \mu} c_m^{\nu} + ie \sum_{ \scr nml \atop \scr \mu 
\nu \ga} c_n^{\da \mu} \left(
t_{nl}^{\mu \ga} \vL_{lm}^{\ga \nu} - \vL_{nl}^{\mu \ga} t_{lm}^{\ga \nu}
\right) c_m^{\nu} - \frac{e^2}{m} \sum_{\scr nm \atop \scr \mu \nu}
\vpot_{n m}^{\mu \nu} c_n^{\da \mu} c_m^{\nu}
\nonumber \\
&=&ie \sum_{\scr nm \atop \scr \mu \nu } c_n^{\da \mu} 
c_m^{\nu} \la n \mu | [ \ham_0(\vpot) , \vr ] | m \nu \ra.
\eea
\end{widetext}
$\ham_0(\vpot) = (\vp - e \vpot)^2/(2m) + V(r)$.
This is just equation (2) expressed in the Wannier basis. The charge current 
is related to the electronic polarization operator~\cite{king}
\[
\pol = e \sum_{\scr nm \atop \scr \mu \nu} c_n^{\da \mu} c_m^{\nu} 
\la n \mu | \vr | m \nu \ra
\]
by $\prl \pol / \prl t = \vj$. The change in polarization $ \Delta \pol$ 
(which is a well defined and measurable bulk quantity, rather than 
polarization 
itself) between an initial and a final state of a sample is the integrated 
current flowing through the sample during an adiabatic transformation 
connecting the two states.~\cite{resta}

Theoretical models of the tight-binding type are effective low 
energy models described in terms of those bands which are close to the Fermi 
surface.~\cite{harrison} 
The question, which is non-trivial and which is still debated, is what should 
be the form of the current for such low energy models.
The low energy Hamiltonian is obtained by eliminating or integrating out the
degrees of freedom corresponding to the high energy bands. This is easily 
formulated in the functional integral language and the procedure generates
many interaction terms that are not present in the original action. In a 
Hamiltonian formulation this is equivalent to making a canonical 
transformation to decouple the low energy and the high energy 
sectors.~\cite{eskes}
That is, given a full many body Hamiltonian $\ham$, 
we perform unitary transformation $U$ such that $U \ham U^{-1}$ is diagonal 
(for a system of interacting particles, in general, this can be done only 
approximately), and then consider only $P U \ham U^{-1} P$, where $P$ is the 
operator projecting on the low energy bands. 
To obtain the expression for the current in the low energy sector one has to 
perform the same canonical transformation used to transform the original 
Hamiltonian into the effective Hamiltonian on the operator representing the 
current. In other words, we first calculate the current (say, $ \vJ$) for the 
full theory (using the symmetry of the full 
theory), make the same unitary transformation and then project the current on 
the low energy sector of interest. 
The exact low energy current is then given by $P U \vJ U^{-1} P$. 
This method of calculating the current for the low energy theory is motivated
by renormalization group ideas. But, to implement this  in practice is 
usually a formidable task. However, if we 
consider a system of non-interacting electrons (in a periodic potential) with 
a subset $M$ of bands that defines the low energy subspace, the low energy 
current is obtained by projecting the full current in eqn. (5) on the low 
energy subspace. This is given by $P \vj P$, where $P = {\displaystyle 
\sum_{\scr n,\mu \in M}} | n \mu \ra \la n \mu |$ 
is the projection operator. We note that the calculation of the exact current
requires knowledge of the matrix elements of the position operator in addition
to that of $\ham_0$ (the tight-binding parameters).~\cite{ram-mohan}

Sometimes, to avoid calculating the matrix elements of the 
position operator, one makes the approximation known as Peierls substitution.
 There are two types of 
approximations involved with this procedure. First, terms involving the 
connection coefficients are dropped out, and one considers an approximate 
gauge transformation given by $\dl c_n^{\mu} = i \ph(\vR_n) c_n^{\mu}$ and 
$\dl c_n^{\da \mu} = -i \ph(\vR_n) c_n^{\da \mu}$. 
Putting the connection coefficients to zero is equivalent to the 
approximation $ \la n \mu | \vr | m \nu \ra \approx \vR_n \dl_{nm} \dl_{\mu 
\nu}$ for the matrix elements of the position operator, and 
$ \la n \mu | \vp | m \nu \ra = i m  \la n \mu | [\ham_0, \vr] | m \nu \ra
\approx i m (\vR_m - \vR_n) t_{nm}^{\mu \nu}$ for the matrix elements of the 
momentum operator.
Second, with this 
approximate gauge transformation, the variation from the interaction term is
non-zero (though, as already noted, it is zero for the exact gauge 
transformation). However, contribution to the current from the interaction 
term is neglected. 
It will be further assumed that the vector potential is constant, i.e., 
$\vpot_{nm}^{\mu \nu} = \vpot \dl_{nm} \dl_{\mu \nu}$.
With these simplifications the approximate current 
($\vj_P$) is given by
\begin{widetext}
\ben
%\lefteqn{
\vj_P = ie \sum_{\scr nm \atop \scr \mu \nu \in M}  
(\vR_m - \vR_n ) t_{nm}^{\mu \nu} c_n^{\da \mu} c_m^{\nu}
%} \nonumber \\ & &
+ e^2 \sum_{\scr nm \atop \scr \mu \nu \in M}(\vR_m - \vR_n ) \left(
(\vR_m - \vR_n ) \cdot \vpot \right) t_{nm}^{\mu \nu} c_n^{\da \mu} c_m^{\nu}
.
\een
\end{widetext}
The second term is the approximate diamagnetic contribution.
The usefulness of $\vj_P$ lies in the fact that it can be calculated from
the tight-binding parameters alone.

The construction of the Peierls current in terms of the atomic orbitals is 
\emph{a priori} not obvious for the case when there is more than one atom 
per unit cell. It is worthwhile to clarify this issue here. We will denote the 
atomic wavefunctions by $| \al \ta \vR_n \ra $, where $\al$ is a symmetry 
index, 
$\vR_n$ is the lattice position of a unit cell, and $\vR_{\ta}$ is the position
of the atom $\ta$ within a unit cell. It is desirable to define the Bloch basis
wavefunctions by $ | \al \ta \vk \ra  = \frac{1}{\sqrt{N}} \sum_{\vR_n} 
e^{-i \vk 
\cdot (\vR_n + \vR_{\ta} )} | \al \ta \vR_n \ra $, though the phase factor 
$e^{-i \vk \cdot \vR_{\ta}}$ is quite innocuous for the definition of the 
Hamiltonian matrix $\ham (\vk)_{\al_1 \ta_1 ; \al_2 \ta_2}$ and for the 
subsequent calculation of the energy bands. The question, whether to keep the
phase factor or not, is however important for the definition of the Peirls 
current $\vj_P (\vk)_{\al_1 \ta_1 ; \al_2 \ta_2} = \frac{\prl}{\prl \vk}
\ham (\vk)_{\al_1 \ta_1 ; \al_2 \ta_2}$. It is easy to verify that, with the 
above definition of the Bloch basis, one gets the same form for the Peierls 
current if one considers a lattice with one atom per unit cell (for which case
the definition of the Peierls current is unambiguous), and compare it with the
same lattice with its period doubled (and therefore now with two identical
atoms per unit cell).

We will examine the behaviour of the exact current and the approximate one
under infinitesimal unitary transformation $U_{nm}^{\mu \nu} = \dl_{nm} 
\dl_{\mu \nu} + W_{nm}^{\mu \nu}$ (where $W$ is antihermitian) of the Wannier
functions defined by $|n \mu \ra \rightarrow \sum_{m \nu} U_{mn}^{\nu \mu} 
| m \nu \ra$. The variation of a matrix element $(\vj)_{nm}^{\mu \nu} = ie
 \la n \mu | [ \ham_0 (\vpot), \vr ] | m \nu \ra $ of the exact current is 
given by
\ben
(\vj)_{nm}^{\mu \nu} \rightarrow (\vj)_{nm}^{\mu \nu} + \sum_{k, \ga}
\left\{ (\vj)_{nk}^{\mu \ga} W_{km}^{\ga \nu} - W_{nk}^{\mu \ga} (\vj)_{km}
^{\ga \nu} \right\}.
\een
This is the usual transformation of matrix elements of operators that remain 
invariant 
under unitary transformation. In fact, the paramagnetic and the diamagnetic 
parts of the operator $\vj$ are separately invariant.
The behaviour of $\vj_P$ is however different.
The variation of $(\vj_P)_{nm}^{\mu \nu} = ie (\vR_m - \vR_n ) t_{nm}^{\mu 
\nu}+ e^2 (\vR_m - \vR_n)((\vR_m - \vR_n) \cdot A) t_{nm}^{\mu \nu}$ 
is given by
\begin{widetext}
\bea
(\vj_P)_{nm}^{\mu \nu} \rightarrow && (\vj_P)_{nm}^{\mu \nu} + \sum_{k, \ga}
\left\{ (\vj_P)_{nk}^{\mu \ga} W_{km}^{\ga \nu} - W_{nk}^{\mu \ga} 
(\vj_P)_{km}^{\ga \nu} \right\} 
%\nonumber \\ &&
+ ie \sum_{k, \ga} \left( \vR_m - \vR_k 
\right) t_{nk}^{\mu \ga} W_{km}^{\ga \nu} 
- ie \sum_{k, \ga} 
\left( \vR_k - \vR_n \right) W_{nk}^{\mu \ga} t_{km}^{\ga \nu}
\nonumber \\ &&
+ e^2 \sum_{k, \ga} \left\{ \left( \vR_k - \vR_n \right) \left( \left( \vR_m 
- \vR_k \right)
\cdot \vpot \right) + \left( \vR_m - \vR_k \right) \left( \left( \vR_m - 
\vR_n \right) \right) \right\} t_{nk}^{\mu \nu} W_{km}^{\ga \nu}
\nonumber \\ &&
- e^2 \sum_{k, \ga} \left\{ \left( \vR_m - \vR_k \right) \left( \left( \vR_k 
- \vR_n \right)
\cdot \vpot \right) + \left( \vR_k - \vR_n \right) \left( \left( \vR_m - 
\vR_n \right) \cdot \vpot \right) \right\} W_{nk}^{\mu \ga} t_{km}^{\ga \nu}.
\eea
\end{widetext}
The paramagnetic and the diamagnetic parts of $\vj_P$ are both basis 
dependent operators.

The basis dependence of $\vj_P$ raises the practical question as to what 
basis one should choose while making the Peierls construction. For example,
there have been efforts to calculate polarization properties, like effective 
charges of semiconductors, using the empirical tight-binding 
theory.~\cite{bennetto} In this 
scheme a natural approximation is the ``diagonal'' ansatz which assumes that 
the position operator is diagonal in the tight-binding basis with expectation
values equal to the atomic positions. This is equivalent to a Peierls 
substitution, and the polarization calculated with this ansatz is related to 
Peierls current $\vj_P$. The effective charges calculated in this procedure
depends on the choice of the underlying Wannier basis. In order to improve 
the results one should first make an appropriate choice of a basis.
One possibility
is to use the basis of the ``maximally localized'' Wannier functions
that was introduced by Marzari and Vanderbilt.~\cite{marzari} This is 
obtained by minimizing a functional which measures the spread of the 
Wannier functions. Intuitively, it seems plausible that the approximation in 
which the connection coefficients are
neglected, will work better in a basis where the Wannier functions are more 
localized. A second possibility, suggested by Millis,~\cite{millis} is 
to choose that basis in which the charge stiffness calculated using the 
Peierls current will be closest to the one obtained from band theory. 
We note that this criteria is already satisfied by the Bloch basis in which 
the effective one-electron Hamiltonian is diagonal in the band indices. This 
can be seen easily in the following manner. We consider the scenario of band 
theory where electrons are in an effective periodic potential. Let $\ep_{\vk
\mu}$ denote the single particle energy levels. It can be shown that the 
charge stiffness is given by $D_{\al \be} = \sum_{\vk \mu} f(\ep_{\vk \mu})
(\prl^2 \ep_{\vk \mu}/(\prl k_{\al} \prl k_{\be} ))$.~\cite{ash} Here 
$f(\ep)$ is the Fermi function and $\al$, 
$\be$ denote spatial directions. The Peierls current constructed in the 
Bloch basis does not have any interband term since the basis is already 
diagonal in the band indices. The paramagnetic part of the current is given
by $ (j_{P})_{{\rm para}, \al} = \sum_{\vk \mu} (\prl \ep_{\vk \mu}/ 
\prl k_{\al}) c_{\vk}^{\da \mu} c_{\vk}^{\mu}$. Since the paramagnetic part 
has no interband matrix element, it does not contribute to the charge 
stiffness. The diamagnetic part, given by $(j_{P})_{{\rm dia}, \al} = -
\sum_{\vk \mu \be} (\prl^2 \ep_{\vk \mu}/ 
(\prl k_{\al} \prl k_{\be})) A_{\be}c_{\vk}^{\da \mu} c_{\vk}^{\mu}$, gives 
a charge stiffness exactly equal to that obtained from band theory. It is 
possible, though, that there are other bases which satisfy this criteria.

In 
passing we note that if the matrix elements of the exact current $\vj$ are
known by some means, say, from first principles calculation, then it is 
possible to define the functional
\ben
\Om = \sum_{\scr nm \atop{\scr \mu \nu \in M}}
\la n \mu | \vj - \vj_P | m \nu \ra \cdot \la m \nu | \vj - \vj_P | n \mu 
\ra
\een
and choose the basis which minimizes $\Om$, and thereby the difference between
the exact current and the approximate one. 
Using eqns. [7] and [8] we can calculate the variation of $\Om$ under 
infinitesimal unitary transformation. The gradient, defined as $G_{nm}
^{\mu \nu} = d \Om / dW_{nm}^{\mu \nu}$, is given by
\begin{widetext}
\bea
G_{nm}^{\mu \nu} &=& 
\left( \vR_m - \vR_n \right) \cdot \la n \mu | [ \ham_0 , (\vj - \vj_P )]
| m \nu \ra 
%\nonumber \\ &&
+ ie \left( (\vR_m \cdot \vpot) \vR_m - (\vR_n \cdot \vpot)
\vR_n \right) \cdot \la n \mu | \{ \ham_0 , (\vj - \vj_P) \} | m \nu \ra
\nonumber \\ &-&
ie \sum_{k \ga} \left\{ (\vR_m - \vR_n)(\vR_k \cdot \vpot) + ((\vR_m - \vR_n
) \cdot \vpot) \vR_k \right \} 
\nonumber \\ &&
\cdot \left \{ \la n \mu | \vj - \vj_P |
k \ga \ra \la k \ga | \ham_0 | m \nu \ra + \la n \mu | \ham_0 | k \ga \ra 
\la k \ga | \vj - \vj_P | m \nu \ra \right\}
\eea
\end{widetext}
The optimum basis is the one for which the gradient vanishes. The choice of 
basis will depend on the vector potential, but the physical quantities 
calculated in that basis will not. 
In general, this criteria will 
give a basis which is different from that of the ``maximally localized'' 
Wannier functions.
The above method of choosing an appropriate basis is not very useful for doing 
charge 
transport calculations because to define the method one needs to know the 
matrix elements of the exact current, knowing which makes the Peierls 
construction redundant. However, one can use this optimization procedure for 
doing thermal transport calculation. As we will see in the next section, the
matrix elements of the exact thermal current are quite complicated, and a 
Peierls formulation of the thermal current is desirable (in some suitable 
basis). The rationale for our suggestion is that the basis which optimizes
the Peierls construction for electric transport will be a good basis for 
doing the Peierls construction for thermal transport as well.

\section{Thermal Current}
In field theory the energy current (which is same as the thermal current, 
except for the latter the single particle energies are measured from the 
chemical potential) is determined from the invariance of the action under the 
transformation of time $t \rightarrow t - \ph(\vr,t)$. This shifts the field 
operators by $\dl \ps = \dps \ph$, and $\dl \psd = \dpsd \ph$. From the 
variation of the action defined in equation (1), the energy current ($\vj_E$)
is given by
\begin{widetext}
\bea
\vj_E = && - \frac{1}{2m} \int d^4 r \left\{ \dpsd \grad \ps
+ \grad \psd \dps \right\} + \frac{1}{4} \int d^3 \vr_1 \int d^3 \vr_2
\left( \vr_2 - \vr_1 \right) 
\nonumber \\
&&
U(r_1 - r_2) \left\{ \dpsd(\vr_1) \rh(\vr_2) \ps(\vr_1) 
- \psd(\vr_1) \dot{\rh}(\vr_2) \ps(\vr_1) + \psd(\vr_1) \rh(\vr_2) \dps(\vr_1)
\right\}.
\eea
\end{widetext}
Here $\rh(\vr) = \psd(\vr) \ps(\vr)$, and $U(r)$ is the two-particle 
interaction energy (Coulomb potential, in our case). The second term above, 
which is formally quartic in the field operators, is the contribution to 
energy current from the non-local (in space) interaction. This term was missed
by Langer,~\cite{langer} but noted in a different context by Jonson and 
Mahan.~\cite{jonson} More recently, it has been discussed by Moreno and 
Coleman.~\cite{moreno}

We have discussed in the previous section that for an effective low-energy
model any current is obtained correctly by projecting the current for the 
full theory (where both high and low energy degrees of freedom are present)
on the low-energy bands. To implement this for the energy current one has
to consider variations of the Wannier operators 
$
\dl c_i^{\mu} = \ph(\vR_i) \cd_i^{\mu} + \grad \ph \sum_{j, \nu} \vL_{ij}^
{\mu \nu} \cd_j^{\nu} 
$
and
$
\dl c_i^{\da \mu} = \ph(\vR_i) \cd_i^{\da \mu} + \grad \ph \sum_{j, \nu} 
\cd_j^{\da \nu} \vL_{ji}^{\nu \mu}
$
under translation of time. If we ignore the terms with the connection 
coefficients, we get an approximate current which is equivalent to a Peierls
substitution. The same approximate current can be derived from the low-energy
effective Hamiltonian using the equations of motion.~\cite{mahan} Although we
are emphasizing the importance of the exact low-energy current, in practice,
calculating the exact thermal current is fairly complicated. Therefore, we 
will restrict the derivation to that of a Peierls type of energy current for a 
generalized Hubbard model described by the Hamiltonian
\ben
\ham =
\sum_{\scr ij \atop {\scr \mu \nu, \si }} t_{ij}^{\mu \nu} 
c_{i \si}^{\da \mu}c_{j \si}^{\nu} + \sum_{\scr ij \atop {\scr \mu \nu,
\si \sip}} V_{ij,\si \sip}^{\mu \nu} n_{i \si}^{\mu} n_{j \sip}^{\nu},
\een
using the equation of motion technique. 
Here $n_{i \si}^{\mu} = c_{i \si}^{\da \mu}c_{i \si}^{\mu}$.
The local energy density ($h_i$) is 
given by
\bea
h_i &=& 
\frac{1}{2} \sum_{\scr j \atop{\scr \mu \nu, \si}} \left(
t_{ij}^{\mu \nu} c_{i \si}^{\da \mu} c_{j \si}^{\nu} + t_{ji}^{\nu \mu} 
c_{j \si}^{\da \nu} c_{i \si}^{\mu} \right) 
\nonumber \\
&+& 
\frac{1}{2} 
\sum_{\scr j \atop{\scr \mu \nu, \si \sip}} \left( V_{ij, \si \sip}
^{\mu \nu} n_{i \si}^{\mu} n_{j \sip}^{\nu} + V_{ji, \sip \si}^{\nu \mu} 
n_{j \sip}^{\nu} n_{i \si}^{\mu} \right). \nonumber
\eea
It can be shown that
\begin{widetext}
\bea
\hd_i &=& \frac{1}{2} \sum_{\scr j \atop{\scr \mu \nu, \si}} 
\left\{ t_{ij}^{\mu \nu} \left(c_{i \si}^{\da \mu} \cd_{j \si}^{\nu} -
\cd_{i \si}^{\da \mu} c_{j \si}^{\nu} \right) + t_{ji}^{\nu \mu} \left(
\cd_{j \si}^{\da \nu} c_{i \si}^{\mu}- c_{j \si}^{\da \nu} \cd_{i \si}^{\mu}
\right) \right\} 
\nonumber \\
&+& \frac{1}{2} \sum_{\scr j \atop{\scr \mu \nu, \si \sip}}
V_{ij, \si \sip}^{\mu \nu} \left( -\cd_{i \si}^{\da \mu} c_{j \sip}^{\da \nu}
c_{j \sip}^{\nu} c_{i \si}^{\mu} + c_{i \si}^{\da \mu} \cd_{j \sip}^{\da \nu}
c_{j \sip}^{\nu} c_{i \si}^{\mu}  
+c_{i \si}^{\da \mu} c_{j \sip}^{\da \nu}
\cd_{j \sip}^{\nu} c_{i \si}^{\mu} - c_{i \si}^{\da \mu} c_{j \sip}^{\da \nu}
c_{j \sip}^{\nu} \cd_{i \si}^{\mu} \right) 
\nonumber \\
&+& \frac{1}{2} \sum_{\scr j \atop{\scr \mu \nu, \si \sip}}
V_{ji, \sip \si}^{\nu \mu} \left( 
\cd_{j \sip}^{\da \nu} c_{i \si}^{\da \mu} c_{i \si}^{\mu} c_{j \sip}^{\nu} -
c_{j \sip}^{\da \nu} \cd_{i \si}^{\da \mu} c_{i \si}^{\mu} c_{j \sip}^{\nu} -
c_{j \sip}^{\da \nu} c_{i \si}^{\da \mu} \cd_{i \si}^{\mu} c_{j \sip}^{\nu} +
c_{j \sip}^{\da \nu} c_{i \si}^{\da \mu} c_{i \si}^{\mu} \cd_{j \sip}^{\nu} 
\right),
\eea
\end{widetext}
where $\dot{\op} = i [\ham, \op]$. The energy current ($ \vj_E $) is related 
to the energy density by the continuity equation 
$
\hd_i + \grad \cdot \vj_E(i) = 0.
$
We define $h(\vq)
= \sum_{i} e^{- i \vq \cdot \vR_i} h_i$, and similarly $\vj_E(\vq)$.
The Fourier transform of the Wannier operators are defined by 
$
c_{\vk \si}^{\mu} = \frac{1}{\sqrt{N}} \sum_{i} e^{-i \vk \cdot \vR_i} 
c_{i \si}^{\mu}, 
$
and similarly for $c_{\vk \si}^{\da \mu}$. Here $N$ is the size of the 
lattice.
Comparing with the continuity equation we get the energy current
\bea
\vj_E &=& \frac{i}{2} \sum_{\scr \vk \atop{\scr \mu \nu, \si}}
\grad_{\vk} \ep_{\vk}^{\mu \nu} \left( c_{\vk,\si}^{\da \mu} \cd_{\vk,\si}
^{\nu} - \cd_{\vk,\si}^{\da \mu} c_{\vk,\si}^{\nu} \right) 
\nonumber \\
&+& \frac{i}{2} \sum_{\scr \vk \vkp \atop{\scr \mu \nu, \si \sip}}
\grad_{\vk} V_{\vk, \si \sip}^{\mu \nu}  \left( c_{\vkp, \si}^{\da \mu}
\nd_{\vk, \sip}^{\nu} c_{\vkp - \vk, \si}^{\mu} \right.
\nonumber \\ 
&-& \left.
\cd_{\vkp, \si}^{\da \mu}
n_{\vk, \sip}^{\nu} c_{\vkp - \vk, \si}^{\mu} - c_{\vkp, \si}^{\da \mu}
n_{\vk, \sip}^{\nu} \cd_{\vkp - \vk, \si}^{\mu} \right),
\eea
where $ n_{\vk, \si}^{\mu} = \sum_{\vkp} c_{\vkp, \si}^{\da \mu}
c_{\vkp + \vk, \si}^{\mu} $.
The first two terms (the quadratic part) in the above eqn. are contributions 
to the energy current from the electron hopping and from the local part of the
interactions. The last three terms (the quartic part) are additional 
contributions to energy flow from the long range interactions. Moreno and 
Coleman~\cite{moreno} have calculated the quartic part using Noether's theorem
for classical fields, and their result is 
$ 
\frac{i}{2} {\displaystyle \sum_{\scr \vk, \mu \nu, \si \sip}}
\grad_{\vk} V_{\vk, \si \sip}^{\mu \nu}  \left( n_{-\vk, \si}^{\mu} 
\nd_{\vk, \sip}^{\nu} - \nd_{-\vk, \si}^{\mu} n_{\vk, \sip}^{\nu} \right).
$
We want to argue that this result is incorrect.
We note that for classical fields the issue of correct arrangement of 
operators is not present. Indeed, if we could commute the third operator with 
the second in each of the last three terms of eqn. (14) we would get the 
result derived in Ref. (15). However such commutation will generate an 
additional term 
$
{\displaystyle \sum_{\scr \vk \vkp, \mu \nu, \si}}
\grad_{\vk} V_{\vk, \si \si}^{\mu \nu} \ep_{\vkp-\vk}^{\mu \nu} c_{\vkp,\si}
^{\da \mu} c_{\vkp, \si}^{\nu}.
$
Thus, proper arrangement of operators is important to get the correct form of
the energy current, which is naturally captured in an equation of motion 
technique but not while using Noether's theorem for classical fields.

The heat current ($ \vj_Q $) is related to the energy current by 
$
\vj_Q = \vj_E - \mu \vj,
$
where $\mu$ is the chemical potential.~\cite{mahan} The chemical potential 
enters only to
shift the single particle energies, i.e., right hand side of eqn. (14) gives
the heat current with the re-definition $\dot{\op} = i [\ham - \mu \pno, 
\op]$, where $\pno$ is the total particle operator.

\section{Transport Coefficients}

\begin{figure}[tbp]
\includegraphics[width=7cm, height=9cm, trim=170 270 200 140]{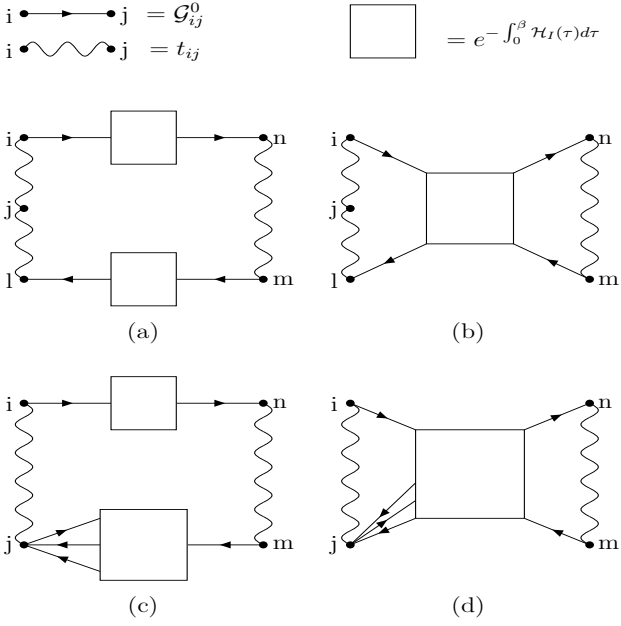}
\caption{Diagrams in configuration space for thermoelectric power. 
$\ham_I$ is the interaction term. In (a) and 
(b) the thermal current is a two-point vertex, while in (c) and (d) it is 
a four-point vertex. In the limit of infinite $d$ contribution from (b) and 
(d) can be neglected.}
\end{figure}

In this section we will examine in detail the derivation of the correlation
functions of the current operators. We will consider only the Peierls type of 
(charge and thermal) currents to keep things analytically tractable. 
In Kubo formalism the correlation functions are related to the corresponding 
response functions
(the transport coefficients). In the framework of DMFT~\cite{gabi}
it is possible to derive exact expressions for the transport 
coefficients. The
essential simplification in the limit of infinite dimensions ($d$) is that 
the self energy and the vertex terms are local. For the single-band Hubbard 
model, defined by the Hamiltonian
\[
\ham = \sum_{\la ij \ra, \si} \left( t_{ij} c_{i, \si}^{\da} c_{j, \si}
 + {\rm h.c.} \right) + U \sum_i n_{i, \uparrow} n_{i, \downarrow},
\]
we will 
demonstrate that this allows the 
correlation functions to be factorized into products of single particle 
Green's functions and their time derivatives. The terms that are ignored by 
such 
factorization are $\ord(1/d)$ smaller and can be neglected in the limit of 
infinite $d$. Using a slightly different approach, the expressions for the 
transport coefficients for the Falikov-Kimball model have been derived 
recently.~\cite{freericks}

The correlation functions of the current operators are defined as~\cite{mahan}
\ben
L_{ab}(i \om_n) = \frac{1}{\be i \om_n V} \int_0^{\be} d \ta e^{i \om_n \ta}
\la T_{\ta} \vj_a(\ta) \vj_b(0) \ra,
\een
where $a, b = (1,2)$, and $\vj_1 = \vj$ is the charge current and $\vj_2
= \vj_Q$ is the heat current. Here $V$ is the volume of the system, 
$\be = 1/k_B T$ is inverse temperature, and $i \om_n$ is bosonic Matsubara 
frequency. The transport coefficients (that enter the formula for DC 
conductivity, thermoelectric power and thermal conductivity) are given by,
\ben
L_{ab} = \lim_{\om \rightarrow 0} {\rm Im} L_{ab}(i \om_n \rightarrow 
\om + i \dl).
\een
For the single band Hubbard model the charge current is given by,
\ben
\vj = e \sum_{\vk, \si} \vv_{\vk} c_{\vk, \si}^{\da} c_{\vk, \si}
=e \sum_{\scr \la ij \ra \atop \scr \si} i \left( \vR_j - \vR_i \right)
t_{ij} c_{i, \si}^{\da} c_{j, \si},
\een
and the heat current is given by
\bea
\vj_Q &=& \frac{i}{2} \sum_{\vk, \si} \vv_{\vk} \left( c_{\vk, \si}^{\da} 
\cd_{\vk, \si} - \cd_{\vk, \si}^{\da} c_{\vk, \si} \right)
\nonumber \\
&=& \frac{1}{2} \sum_{\scr \la ij \ra \atop \scr \si} \left( \vR_i 
- \vR_j \right) t_{ij} \left( c_{i, \si}^{\da} \cd_{j, \si} - 
\cd_{i, \si}^{\da} c_{j, \si} \right).
\eea
Here $ \vv_{\vk}= \grad_{\vk} \ep_{\vk}$ is the electron velocity.
Since the interaction is purely local, there is no contribution from the 
long range interactions.

The derivation of $L_{11}$ is discussed extensively in the literature on 
DMFT.~\cite{pruschke,schweitzer}
In infinite $d$ the particle-hole vertex becomes momentum 
independent,~\cite{khurana} and  
the dressed correlation function becomes equal to the bare one. This implies
the correlation function can be factorized into a product of single particle 
Green's functions, i.e.,
$
\la T_{\ta} \vj(\ta) \vj(0) \ra = - \frac{e^2}{d} \sum_{\vk, \si} v_{\vk}^2
\mg_{\si}(\vk, \ta) \mg_{\si}(\vk, -\ta),  
$
where $\mg_{\si}(\vk, \ta) = - \la T_{\ta} c_{\vk, \si}(\ta) c_{\vk, \si}
^{\da}(0) \ra$ is the fermionic Matsubara Green's function. We define the 
Fourier transform 
$
\mg_{\si}(\vk, \ta) = \frac{1}{\be} \sum_n e^{-i \om_n \ta} \mg_{\si}(\vk, 
i \om_n),
$
in terms of which
\bea
\lefteqn{
L_{11}(i \om_n) = 
} \nonumber \\ &&
- \left( \frac{e^2}{\be i \om_n d V}
\right) \sum_{\vk, \si, i p_n} v_{\vk}^2 \frac{1}{\be}
\mg_{\si}(\vk, i \om_n + i p_n) \mg_{\si}(\vk, i p_n).
\nonumber
\eea
$\mg_{\si}(\vk, z)$ has a possible branch cut at $z= \ep$ and 
$\mg_{\si}(\vk, z + i \om_n)$ has one at $z= \ep - i \om_n$.~\cite{mahan}
Following Mahan~\cite{jonson,mahan} one can show
\bea
\lefteqn{
\frac{1}{\be} \sum_{i p_n} 
\mg_{\si}(\vk, i \om_n + i p_n) \mg_{\si}(\vk, i p_n)
= 
} \nonumber \\ &&
\int_{-\infty}^{\infty} \frac{d \ep}{2 \pi } n_F(\ep) A_{\si}(\vk, \ep)
\left[ \mg_{\si}(\vk, \ep + i \om_n) + \mg_{\si}(\vk, \ep - i \om_n) \right],
\nonumber
\eea
where $A_{\si}(\vk, \ep) = - 2 {\rm Im} G^R_{\si}(\vk, \ep)$ is the spectral 
function and $n_F(\ep)$ is the Fermi function. After analytic continuation
$i \om_n \rightarrow \om + i \dl$, and after taking the static limit we get
\ben
L_{11} = \frac{e^2}{2d \be V} \sum_{\vk, \si} v_{\vk}^2 \int_{-\infty}
^{\infty} \frac{d \ep}{2 \pi }\left( - \frac{\prl n_F(\ep)}{\prl \ep} \right) 
A^2_{\si}(\vk, \ep).
\een

The derivation of $L_{21}$ is more involved, and is not well discussed in 
the literature. Since the heat current has a part which is a four-point 
vertex, \emph{a priori} it is not clear whether a factorization of the 
correlation 
function into products of single particle Green's functions and their time 
derivatives is possible. We have 
$
\cd_{i, \si} = -i \sum_l t_{il} c_{l, \si} - i U c_{i, \si} n_{i, \sib}
+ i \mu c_{i, \si}
$
(and similarly for $\cd_{i, \si}^{\da}$). We ignore the term with the chemical
potential for the time being (the result remains unchanged). Due to the first 
term the heat current is a two-point vertex, and the corresponding diagrams 
for $L_{21}$ are of the type (a) and (b) of Fig. 1. The heat current is a 
four-point vertex due to the second term. The corresponding diagrams are of 
the type (c) and (d) of Fig. 1.  In the limit of infinite $d$ the scaling of 
the hopping term is 
$t_{ij} = t^{\ast}_{ij}/\sqrt{d}$ (Ref. 4). This implies that 
$\mg_{ij}^0 \sim (1/\sqrt{d})^{|i-j|}$ (Ref. 4). One can show explicitly 
that diagrams (a) and (c) are
$\ord(1/d)$ (and higher), and diagrams (b) and (d) are $\ord(1/d^2)$ (and 
higher). 
In Fig.1, $\ham_I = U \sum_i n_{i, \uparrow} n_{i, \downarrow}$
is the interaction term of the Hubbard Hamiltonian.
In the limit of infinite $d$ the latter drops out, and the 
factorization of the correlation function is possible. In imaginary time
\bea
\lefteqn{
\la T_{\ta} \vj_Q (\ta) \vj (0) \ra \stackrel{d \rightarrow \infty}{=}
} \nonumber \\ &&
\frac{e}{2d} \sum_{\vk, \si} v_{\vk}^2 \left\{ \la T_{\ta} 
\cd_{\vk, \si} (\ta) c_{\vk, \si}^{\da}(0) \ra \la T_{\ta} c_{\vk, \si}(0)
c_{\vk, \si}^{\da}(\ta) \ra + {\rm h.c.} \right\}.
\nonumber 
\eea
Using $\frac{\prl}{\prl \ta} \mg(\ta) = \la T_{\ta} \frac{\prl}{\prl \ta}
c(\ta) c^{\da}(0) \ra - \dl(\ta)$ (in imaginary time), we get
\begin{widetext}
\[
L_{21}(i \om_n) = - \left( \frac{e}{d} \right) \left( \frac{1}{\be i \om_n V}
\right) \sum_{\vk, \si} v_{\vk}^2 \left\{ \frac{1}{\be} \sum_{i p_n}
\left( i p_n + \frac{i \om_n}{2} \right) \mg_{\si}(\vk, ip_n) 
\mg_{\si}(\vk, ip_n + i \om_n) - n_{\vk, \si} \right\}.
\]
\end{widetext}
We drop the second term within braces because it does not contribute to 
${\rm Im} L_{21}(\om + i \dl)$. The rest is evaluated like $L_{11}(i \om_n)$.
It can be shown that
\bea
\lefteqn{ 
\frac{1}{\be} \sum_{i p_n} \left( i p_n + \frac{i \om_n}{2} \right) 
\mg_{\si}(\vk, ip_n) \mg_{\si}(\vk, ip_n + i \om_n) =
} \nonumber \\ &&
\int_{-\infty}^{\infty} \frac{d \ep}{2 \pi } n_F(\ep) A_{\si}(\vk, \ep)
\left[ \left( \ep + \frac{i \om_n}{2} \right)\mg_{\si}(\vk, \ep + i \om_n)
\right. \nonumber \\ && \left.
+ \left(\ep - \frac{i \om_n}{2} \right) \mg_{\si}(\vk, \ep - i \om_n) \right].
\nonumber
\eea
After analytic continuation and taking the static limit we get,
\ben
L_{21} = \frac{e}{2d \be V} \sum_{\vk, \si} v_{\vk}^2 \int_{-\infty}
^{\infty} \frac{d \ep}{2 \pi } \ep \left( - \frac{\prl n_F(\ep)}{\prl \ep} 
\right) A^2_{\si}(\vk, \ep).
\een

The derivation of $L_{22}$ is analogous to that of $L_{21}$. In the limit of 
infinite $d$, $\la T_{\ta} \vj_Q(\ta) \vj_Q \ra$ factorizes into products of
(imaginary) time derivatives of single particle Green's functions (plus terms 
which do not contribute to ${\rm Im} L_{22}(\om)$). As in the case of $L_{11}$
and $L_{21}$, the terms which are dropped out by such factorization are at 
least $\ord(1/d)$ smaller.  In other words,
\bea
\lefteqn{
\la T_{\ta} \vj_Q(\ta) \vj_Q \ra \stackrel{d \rightarrow \infty}{=} 
} \nonumber \\ &&
\frac{1}{4d} \sum_{\vk, \si} v_{\vk}^2 \left\{ \la T_{\ta} \cd_{\vk, \si}(\ta)
c_{\vk, \si}^{\da}(0)\ra \la T_{\ta} c_{\vk, \si}(0) \cd_{\vk, \si}^{\da} 
(\ta) \ra 
\right. \nonumber \\ && \left.
- \la T_{\ta} \ddot{c}_{\vk, \si} (\ta) c_{\vk, \si}^{\da}(0) \ra 
\la T_{\ta} c_{\vk, \si}(0) c_{\vk, \si}^{\da}(\ta) \ra + {\rm h.c.} \right\}.
\nonumber
\eea
With this simplification it can be shown that
\begin{widetext}
\[
L_{22}(i \om_n) = - \left( \frac{1}{d} \right) \left( \frac{1}{\be i \om_n V}
\right) \sum_{\vk, \si} v_{\vk}^2 \left\{ \frac{1}{\be} \sum_{i p_n}
\left( i p_n + \frac{i \om_n}{2} \right)^2 \mg_{\si}(\vk, ip_n) 
\mg_{\si}(\vk, ip_n + i \om_n) + \cdots \right\}.
\]
\end{widetext}
The terms in the ellipses do not contribute to ${\rm Im} L_{22}(\om)$. Finally
we get,
\ben
L_{22} = \frac{e}{2d \be V} \sum_{\vk, \si} v_{\vk}^2 \int_{-\infty}
^{\infty} \frac{d \ep}{2 \pi } \ep^2 \left( - \frac{\prl n_F(\ep)}{\prl \ep} 
\right) A^2_{\si}(\vk, \ep).
\een
We reiterate the observation made in Ref. (8) that the above expressions 
for the transport coefficients are correct for any model with local
interaction (for which Eq. [18] is correct), in infinite dimensions.
 
\section{Conclusion}
The current (charge or thermal) obtained by Peierls 
substitution or by the equation of motion technique is an approximation to the
exact low energy current for an effective tight-binding Hamiltonian. In 
particular, the approximate current is not invariant under a unitary  
transformation of the Wannier basis. We have suggested a simple criteria by 
which one can choose a set of Wannier functions where the difference between
the exact and the approximate current is minimum. The minimization procedure 
is well defined provided the matrix elements of the exact current are known
from first principles calculation. Using the equations of
motion we have derived the thermal current for a very general tight-binding
Hamiltonian, correcting the result of a previous work. Finally, using the 
Peierls
currents, we have established the correctness of known expressions for the 
transport coefficients for the
Hubbard model in infinite $d$. The simplification in the limit of large 
coordination is that the 
current (charge and thermal) correlation functions can be factorized into 
products of single particle Green's functions and their time derivatives.  
These expressions are correct for any model with local interaction and in 
infinite dimensions.

\section{Acknowledgments} 
We thank D. Vanderbilt, I. Souza, G. Palsson, A. J. Millis, and J. K. 
Freericks for useful discussions. This research 
was supported by the Division of Materials Research of the National Science
Foundation under Grant No. DMR-0096462.

\end{document}